\documentclass[onecolumn,preprintnumbers,amsmath,amssymb,groupaddress]{revtex4}

\usepackage{graphicx}
\usepackage{dcolumn}
\usepackage{bm}
\usepackage{CJK}
\usepackage{threeparttable}

\begin{document}
\begin{CJK*}{GBK}{song}

\title{Multi-peak solitons in nonlocal nonlinear system with sine-oscillation response}

\author{Lanhua Zhong$^{a,b}$}
\author{Dalong Dang$^a$}
\author{Wei Li$^a$}
\author{Zhanmei Ren$^a$}
\author{Qi Guo$^a$}\email{guoq@scnu.edu.cn}

\address{$^a$Guangdong Provincial Key Laboratory of Nanophotonic Functional Materials and Devices, South China Normal University, Guangzhou 510631, China}
\address{$^b$Physical Science and Technology School, Lingnan Normal University, Zhanjiang 524048, China}

\date{\today}

\begin{abstract}
The multi-peak solitons and their stability are investigated for the nonlocal nonlinear system with the sine-oscillation response, including both the cases of positive and negative Kerr coefficients. The Hermite-Gaussian-type multi-peak solitons and the ranges of the degree of nonlocality within which the solitons exist are analytically obtained by the variational approach. This is the first time, to our knowledge at least, to discuss the solution existence range of the multi-peak solitons analytically, although approximately. The variational analytical results are confirmed by the numerical ones.
 The stability of the multi-peak solitons are addressed by the linear stability analysis. It is found that the upper thresholds of the peak-number of the stable solitons are five and four for the system with negative and positive Kerr coefficients, respectively.
\end{abstract}


\maketitle
\end{CJK*}
\section{Introduction}
Nonlocality is ubiquitous and of great importance in many physical systems. In optics, a nonlinear medium is considered spatially nonlocal when the nonlinear response of the material to the light wave is determined not only by the wave at that point but also by its vicinity. The spatially nonlocal nonlinearity appears if the interaction between light and matter involves a mechanism such as the diffusion of carriers, reorientation of molecules, heat conduction, etc~\cite{Assanto-book-13, Guo-book-15}. Since the pioneering work done by Snyder and Mitchell\cite{Snyder-science-97},  the research on optical beams in nonlocal nonlinear media has been brought in focus. Some particular properties resulting from nonlocality have been found, such as the long-range interactions between beams~\cite{Rasmussen-pre-05, Rotschild-np-06, Hu-apl-06}, the large phase shift of solitons~\cite{Guo-pre-04, Shou-ol-11, Shou-oc-15},  the supports of complex spatial solitons~\cite{Rotschild-prl-05, Xu-ol-05, Rotschild-ol-06, Skupin-pre-06, Buccoliero-prl-07, Skupin-prl-07, Deng-josab-07, Buccoliero-ol-08, Rotschild-np-08, Dong-pra-10, Song-oe-18} and chaoticons~\cite{Zhong-sr-17, Zhong-pra-19}, etc.

However, most of the researches concentrated on the nonlocal nonlinearity with positive definite and localized response function, such as the phenomenological Gaussian response function~\cite{Guo-pre-04, Buccoliero-prl-07, Deng-josab-07, Song-oe-18}, the logarithmic response function~\cite{Shou-ol-09, Zeng-pra-18} in the lead-glass ~\cite{Rotschild-prl-05, Dong-pra-10, Shou-ol-09, Zeng-pra-18} for negligible loss thermal-nonlinearity~\cite{Ghofraniha-PRL-99}, the exponential-decay response function~\cite{Rasmussen-pre-05, Xu-ol-05, Zhong-sr-17, Zhong-pra-19} and the zeroth-order modified Bessel function~\cite{ Hu-apl-06, Skupin-pre-06} in media with a nonlinearity described by the diffusion-type equation~\cite{Skupin-pre-06}, e.g., the effect of plasma heating on the
propagation of electromagnetic waves~\cite{Litvak-jl-1966}, the orientational nonlinearity of the nematic liquid crystal~\cite{Assanto-book-13, Conti-prl-04}, and the thermal-nonlinearity in the regime of strong absorption~\cite{Ghofraniha-PRL-99}.
Recently, some attention has been paid to the nonlocal nonlinear system with the sine-oscillation response function~\cite{Esbensen-pra-12, Wang-ol-14, Wang-oc-17, Liang-oe-16, Chen-pra-18, Liang-pra-19, Liang-njp-20, Guan-sr-20}, which was first put forward by Nikolov et al in the study of quadratic solitons~\cite{Nikolov-pre-03}. The distinct difference of such a response function from the localized ones lies in that it is periodically oscillatory. The peculiarity of the response function may lead to some novel properties for the nonlocal nonlinear system. For example, the very recent research~\cite{Liang-pra-19, Liang-njp-20} revealed that there appears a transition between self-focusing and self-defocusing in this kind of nonlocal nonlinear media when the degree of nonlocality goes across a critical value.

As it is known, in the nonlocal nonlinear systems with localized response functions, there exist not only the fundamental solitons, but also high order solitons with multi-peak~\cite{Xu-ol-05, Dong-pra-10, Deng-josab-07}. The reason may relate to the fact that the nonlinear refractive index (NRI) is approximate to the parabolic shape on the condition of strong nonlocality, which possesses the Hermite-Gaussian (HG) soliton solutions with multi-peaks. However, the upper thresholds of the peak-number of stable  solitons supported are different in the system with different response function.
For instance, in the nonlocal system with the exponential-decay response, only the multi-peak solitons with the peak-number less than five are stable~\cite{Xu-ol-05, Zhong-sr-17}. While in the system with the Gaussian response, the multi-peak solitons with any peak-number are stable~\cite{Xu-ol-05, Deng-josab-07}.
For the nonlocal nonlinear system with the sine-oscillation response, the fundamental solitons and a kind of dipole soliton have been discussed~\cite{Liang-oe-16, Liang-pra-19, Liang-njp-20}. Here we will study the existence and the stability of the multi-peak solitons in this system by both the analytical and numerical methods.

The paper is organized as follows. In Sec.~\ref{sec2}, the model describing the nonlocal nonlinear system with the sine-oscillation response is introduced. In Sec.~\ref{sec3}, the HG multi-peak solitons and their existence condition are investigated by the variational approach. In Sec.~\ref{sec4}, the numerical multi-peak solitons are acquired, and the relations between several quantities are discussed. In Sec.~\ref{sec5}, the stability of the multi-peak solitons is studied by the linear stability analysing. Section \ref{sec6} gives the conclusion that the variational and numerical results agree with each other, and the upper thresholds of the peak-number of the stable solitons are five and four in the cases with negative and positive Kerr coefficients, respectively.

\section{Model}\label{sec2}
We consider the paraxial propagation of a (1+1)-dimensional optical beam in a nonlocal nonlinear medium  described by the coupled equations~\cite{Liang-oe-16, Liang-pra-19, Liang-njp-20}
\begin{equation}\label{se}
i\frac{\partial q}{\partial z}+\frac{1}{2}\frac{\partial^{2}q}{\partial x^{2}}+q\Delta n=0,
\end{equation}
\begin{equation}\label{nri}
w_{M}^{2}\frac{\partial^{2}\Delta n}{\partial x^{2}}+\Delta n-s|q|^{2}=0,
\end{equation}
in which $q(x,z)$ and $\Delta n(x,z)$ are the dimensionless complex optical field amplitude and the light-induced NRI, respectively, $x$ and $z$ stand for the transverse and longitudinal coordinates,  respectively, and $w_{M}$ is the nonlinear characteristic length of the media, $s(=\pm 1)$ is the sign of the Kerr coefficient.
When the Kerr coefficient is negative ($s=-1$), Eqs.~(\ref{se}) and (\ref{nri}) resemble the model describing the propagation of beams in the nematic liquid crystals~\cite{Rasmussen-pre-05, Xu-ol-05, Zhong-sr-17, Zhong-pra-19}, where the only difference is that the sign before the second term is minus in Eq.~(\ref{nri}).
When the Kerr coefficient is positive ($s=1$), Eqs.~(\ref{se}) and (\ref{nri}) are identical with the second harmonic generation model in media with the quadratic nonlinearity for the stationary solutions~\cite{Buryak-pla-95}. Thus, these equations  can be deemed a reasonable extension of the models describing the real physics~\cite{Liang-pra-19, Liang-njp-20}.

When the boundary of Eq.~(\ref{nri}) satisfies certain conditions~\cite{Liang-pra-19}, the NRI can be expressed as a convolution
\begin{equation}\label{index}
\Delta n(x, z)=sR\otimes|q|^{2}=s\int_{-\infty}^{\infty}R(x-x')|q(x',z)|^{2}\mathrm{d}x',
\end{equation}
 where the symbol $\otimes$ denotes the convolution, the response function $R$ is of the sine-oscillation form
\begin{equation}\label{rf}
R(x)=\frac{1}{2w_{M}}\sin\left (\frac{|x|}{w_{M}}\right ).
\end{equation}
We can find easily that the oscillation period of the above response function is $2\pi w_{M}$. Then Eq.~(\ref{index}) are equivalent to Eq.~(\ref{nri}).

The soliton (stationary) solutions of Eqs.~(\ref{se}) and (\ref{index}) have the form
\begin{equation}\label{ss}
 q(x,z)=u(x)\exp(ibz),
\end{equation}
 in which both the profile function $u$ and the propagation constant $b$ are real. The existence and the shape of the soliton solution depend not only on the sign of $s$, but also on the generalized degree of nonlocality (GDN)~\cite{Liang-oe-16, Liang-pra-19} $\sigma=w_{M}/w$, in which the beam width is defined by the second order moment
\begin{equation}\label{bw}
w=\left (2\int_{-\infty}^{\infty}x^{2}|q|^{2}\mathrm{d}x/\int_{-\infty}^{\infty}|q|^{2}\mathrm{d}x\right )^{1/2}.
\end{equation}
 As stated in previous works~\cite{Liang-pra-19}, there exists the intertransition between the self-focusing and self-defocusing nonlinearity.
 When $s=-1$ and $\sigma>0.82$, or $s=1$ and $\sigma<0.82$, the system exhibits the self-focusing nonlinearity,
 and the fundamental solitons~\cite{Liang-oe-16, Liang-pra-19} and a kind of out-of-phase solitons~\cite{Liang-oe-16} had been found. Here in this paper, we will discuss the bright solitons with multi-peak analytically and numerically for the self-focusing nonlinearity state.

\section{Variational procedure}\label{sec3}
Now we will find the approximate multi-peak solitons analytically by using the variational approach. The Lagrangian density of the system described by Eqs.~(\ref{se}) and (\ref{index}) is~\cite{Anderson-pra-83, Zhong-oc-17}
\begin{equation}\label{density}
l=\frac{i}{2}\left (q^{*}\frac{\partial q}{\partial z}-q\frac{\partial
q^{*}}{\partial z}\right )-\frac{1}{2}\left |\frac{\partial q}{\partial
x}\right |^{2}+\frac{s}{2}|q|^{2}\int^{\infty}_{-\infty}R(x-\xi)|q(\xi,z)|^{2}\mathrm{d}\xi,
\end{equation}
where the superscript $*$ denotes the conjugate complex. We search the multi-peak solitons with the HG shape, so the trial solution $q(x,z)$ for the variational problem is supposed to be~\cite{Deng-josab-07, Zhong-oc-17}
\begin{equation}\label{trial}
q_{n}(x,z)=\frac{A_{n}(z)\exp[i \alpha_{n}(z)]}{(\sqrt{\pi} 2^{n}n!)^{1/2}}H_{n}\left [\frac{\sqrt{2n+1}x}{w_{n}(z)}\right ]\exp\left[-\frac{(2n+1)x^{2}}{2w_{n}^{2}(z)}+ic_{n}(z)x^{2}\right],
\end{equation}
where $H_{n}(x)=(-1)^{n}e^{x^{2}}\mathrm{d}^{n}e^{-x^{2}}/\mathrm{d}x^{n}$ is the $n$th-order ($n=0,1,2,3,...$) Hermite polynomial~\cite{Abramowitz-book-72}, $A_{n}$ and $\alpha_{n}$ are the amplitude and phase of the complex amplitude of the solution, respectively, $c_{n}$ is the phase-front curvature, $w_{n}$ is exactly the beam width given in Eq.~(\ref{bw}). All of $A_{n}$, $\alpha_{n}$, $w_{n}$ and $c_{n}$ are the real functions of $z$, the power $P_{n}=\int_{-\infty}^{\infty}|q_{n}|^{2}\mathrm{d}x=A_{n}^{2}w_{n}/\sqrt{2n+1}$. When $n=0$, $H_{0}(x)=1$, Eq.~(\ref{trial}) degenerates to the Gaussian form with the shape of single peak.

By substituting the trial solution (\ref{trial}) into the Lagrangian density (\ref{density}), using the expression $H_{n+1}(x)-2xH_{n}(x)+2nH_{n-1}(x)=0$ and the orthogonality of the Hermite polynomial, the Lagrangian $L=\int_{-\infty}^{\infty}l\mathrm{d}x$ can be obtained
\begin{equation}\label{lagran}
L=-\frac{w_{n}^{3}A_{n}^{2}}{2\sqrt{2n+1}}\left[\frac{\mathrm{d}c_{n}}{\mathrm{d}z}+\frac{(2n+1)^{2}}{2w_{n}^{4}}+2c_{n}^{2}\right]
-\frac{w_{n}A_{n}^{2}}{\sqrt{2n+1}}\frac{\mathrm{d}\alpha_{n}}{\mathrm{d}z}+\frac{sA_{n}^{4}}{2(\sqrt{\pi}2^{n}n!)^{2}}e_{n}(w_{n},w_{M}),
\end{equation}
where $e_{n}(w_{n},w_{M})=\int_{-\infty}^{\infty}H_{n}^{2}\exp[-(2n+1)x^{2}/w_{n}^{2}]\{R(x)\otimes H_{n}^{2}\exp[-(2n+1)x^{2}/w_{n}^{2}]\}\mathrm{d}x$.
Following the standard procedures of the variational approach~\cite{Anderson-pra-83}, we obtain the second order differential equation for the width of the HG beam (The detailed deviation is given in appendix A.)
\begin{equation}\label{width}
\frac{\mathrm{d}^{2}w_{n}}{\mathrm{d}z^{2}}=\frac{(2n+1)^{2}}{w_{n}^{3}}-\frac{sP_{n}(2n+1)}{(\sqrt{\pi}2^{n}n!)^{2}}N_{n}(w_{n},w_{M}),
\end{equation}
where
\begin{equation}\label{Nn}
N_{n}(w_{n},w_{M})=\frac{2e_{n}(w_{n},w_{M})-f_{n}(w_{n},w_{M})}{w_{n}^{3}},
\end{equation}
and $f_{n}(w_{n},w_{M})=\int_{-\infty}^{\infty}(H_{n+1}^{2}-4n^{2}H_{n-1}^{2})\exp[-(2n+1)x^{2}/w_{n}^{2}]\{R(x)\otimes H_{n}^{2}\exp[-(2n+1)x^{2}/w_{n}^{2}]\}\mathrm{d}x$.

 By comparing Eq.~(\ref{width}) with the Newton's second law in the classical mechanics~\cite{Guo-pre-04}, the right hand side can be viewed as the ``force"
 \begin{equation}\label{force}
  F=\frac{(2n+1)^{2}}{w_{n}^{3}}-\frac{sP_{n}(2n+1)}{(\sqrt{\pi}2^{n}n!)^{2}}N_{n}(w_{n},w_{M})
 \end{equation}
acted on a particle with the unit mass for the one-dimensional motion. Here the beam width $w_{n}$ and the longitudinal coordinate $z$ are equivalent to the spatial and temporal coordinates of the particle, respectively. The ``potential" of the force is $V_{n}=-\int_{w_{n0}}^{w_{n}}F\mathrm{d}w_{n}$. The existence of the stationary solutions, equivalently, the stable equilibrium state of the particle, appears on the minimal point of the potential. Then the two conditions should be met: $\mathrm{d}V_{n}/\mathrm{d}w_{n}(=-F)=0$ and $\mathrm{d}^{2}V_{n}/\mathrm{d}w_{n}^{2}(=-\mathrm{d}F/\mathrm{d}w_{n})>0$.

According to the condition of $\mathrm{d}V_{n}/\mathrm{d}w_{n}=0$ and Eq.~(\ref{force}), we can easily obtain the critical power $P_{nc}=(2n+1)(\sqrt{\pi}2^{n}n!)^{2}/[sw_{n}^{3}N_{n}(w_{n},w_{M})]$, which is the power carried by the $n$-th order HG solitons. The beam is assumed to be incident at its waist, that is, $[\mathrm{d}w_{n}(z)/\mathrm{d}z]|_{z=0}=0$ (the equivalent  particle starts from rest). And the initial ``position" is regarded as $w_{n}(z)|_{z=0}=w_{n0}=1$ for the dimensionless system. Then the nonlinear
characteristic length $w_{M}=\sigma w_{n}=\sigma_{0}|_{w_{n0}=1}$, where the initial GDN $\sigma_{0}=\sigma(z)|_{z=0}$. The beam will keep the width $w_{n}(z)=w_{n0}=1$ unchanged during the propagation, since the equivalent particle at rest will never move when it is acted by the balance force. Hence the  critical power can be rewritten as
\begin{equation}\label{pc}
P_{nc}=\frac{(2n+1)(\sqrt{\pi}2^{n}n!)^{2}}{sN_{n}(1,\sigma_{0})}.
\end{equation}
Therefore, the trail solution Eq.~(\ref{trial}) is of the form of the soliton solution [Eq.~(\ref{ss})]: $q_{n}(x,z)=u_{n}(x)\exp(ib_{n}z)$, where the propagation constant is
$b_{n}=\mathrm{d}\alpha_{n}/\mathrm{d}z=-(2n+1)^{2}/2+3(2n+1)^{2}[2e_{n}-f_{n}/3]/[4N_{n}(1,\sigma_{0})]$, the profile function is
\begin{equation}\label{un}
u_{n}=\left(\frac{P_{nc}\sqrt{2n+1}}{2^{n}n!\sqrt{\pi}}\right)^{1/2}H_{n}(\sqrt{2n+1}x)\exp\left[-\frac{(2n+1)x^{2}}{2}\right].
\end{equation}
Obviously, the profile of the $n$-th order HG soliton possesses $n+1$ peaks.

To consider the condition of $\mathrm{d}^{2}V_{n}/\mathrm{d}w_{n}^{2}>0$, we obtain  firstly
\begin{equation}\label{potent}
\begin{split}
\frac{\mathrm{d}^{2}V_{n}}{\mathrm{d}w_{n}^{2}}=&\frac{(2n+1)^{2}}{N_{n}(1,\sigma_{0})}\left[2f_{n}(1,\sigma_{0})-\frac{\partial f_{n}(1,\sigma_{0})}{\partial w_{n}}\right]\\
=&\frac{2(2n+1)^{2}}{N_{n}(1,\sigma_{0})}\Phi_{n}(1,\sigma_{0}),
\end{split}
\end{equation}
where
\begin{equation}\label{fi}
\begin{split}
\Phi_{n}(1,\sigma_{0})=&\int^{+\infty}_{-\infty}\left[H^{2}_{n+1}-4n^{2}H^{2}_{n-1}+(2n^{2}+2n+1)H^{2}_{n}-4n^{2}(n-1)^{2}H^{2}_{n-2}-\frac{H^{2}_{n+2}}{4}\right]\\
&\exp[-(2n+1)x^{2}]\left\{R(x)\otimes H^{2}_{n}\exp[-(2n+1)x^{2}]\right\}\mathrm{d}x\\
&-\int^{+\infty}_{-\infty}\left(\frac{H^{2}_{n+1}}{4}-n^{2}H^{2}_{n-1}\right)\exp[-(2n+1)x^{2}]\\
&\left\{R(x)\otimes(H^{2}_{n+1}-4n^{2}H^{2}_{n-1})\exp[-(2n+1)x^{2}]\right\}\mathrm{d}x.
\end{split}
\end{equation}
 It is clear from Eq.~(\ref{potent}) that a positive value of $\mathrm{d}^{2}V_{n}/\mathrm{d}w_{n}^{2}$ requires the same signs of $N_{n}$ and $\Phi_{n}$. Additionally, the  power $P_{nc}$ [given by Eq.~(\ref{pc})] is demanded to be positive for a physical system, which requires that the signs of $s$ and $N_{n}$ should also be same. Therefore, for the case of $s=-1$, the inequations of $N_{n}<0$ and $\Phi_{n}<0$ should hold. On the contrary, for the case of $s=1$, the inequations of $N_{n}>0$ and $\Phi_{n}>0$ should hold.
Then we can find the ranges of $\sigma_{0}$ within which the above conditions are satisfied by solving numerically the
algebraic equations of $N_{n}(1,\sigma_{0})=0$ and $\Phi_{n}(1,\sigma_{0})=0$, where the two functions $N_{n}$ and $\Phi_{n}$ are given by Eqs.~(\ref{Nn}) and (\ref{fi}), respectively.
Without loss of generality, the triple-solitons ($n=2$) are exemplified. When $s=-1$, the inequations of $N_{n}(1,\sigma_{0})<0$ and $\Phi_{n}(1,\sigma_{0})<0$ correspond to the ranges that $\sigma_{0}\in (0.12, 0.14)\cup(0.24, 0.30)\cup(1.06, +\infty)$. When $s=1$, the inequations of $N_{n}(1,\sigma_{0})>0$ and $\Phi_{n}(1,\sigma_{0})>0$ correspond to the ranges that $\sigma_{0}\in (0,0.06)\cup(0.09, 0.11)\cup(0.17, 0.21)\cup(0.39, 0.79)$.

\begin{figure}[htbp]
  \centering
  \includegraphics[width=8.5cm]{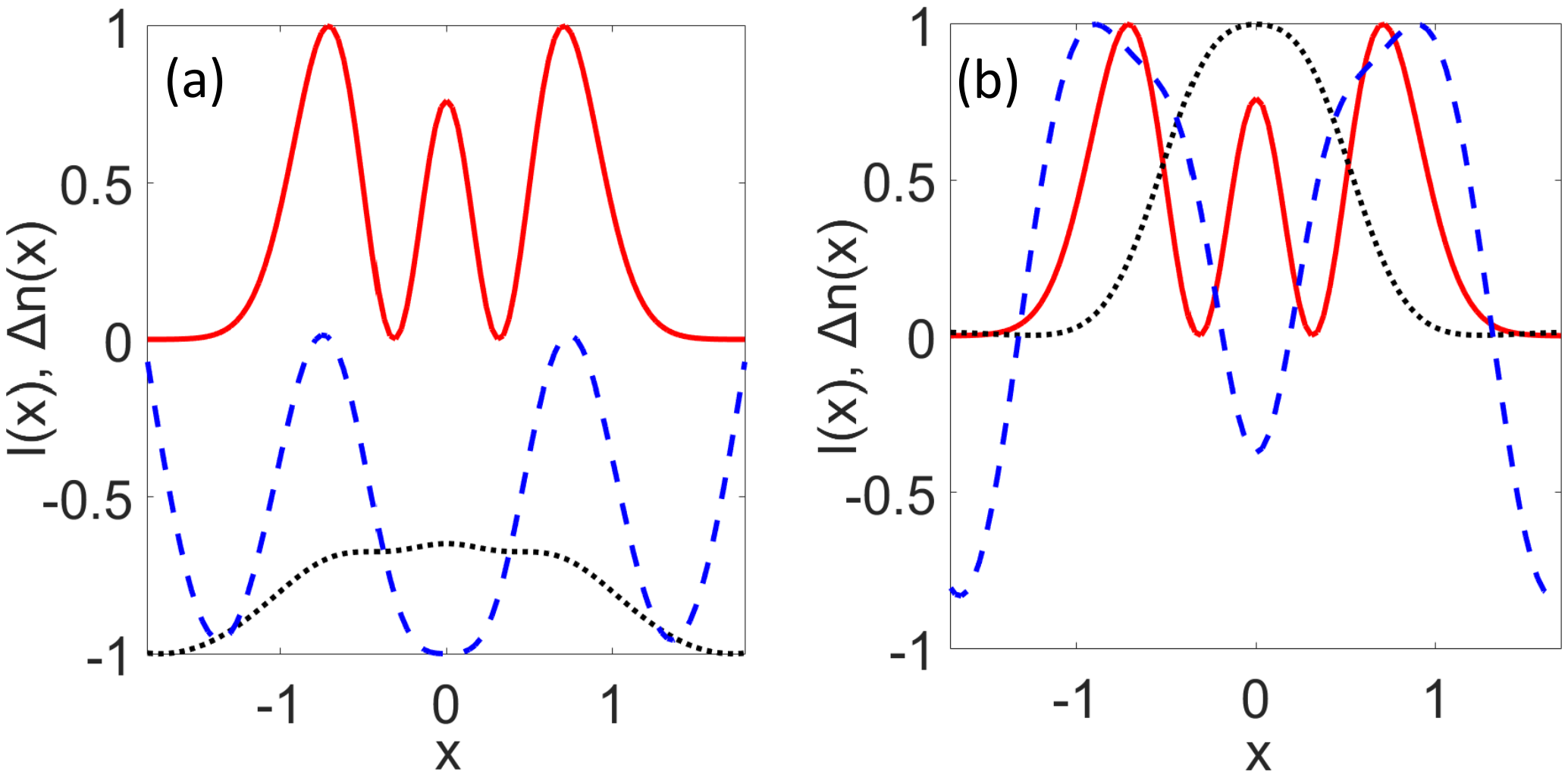}
  \caption{The NRI $\Delta n$ (normalized values) induced by the triple-soliton at different $\sigma_{0}$. (a) $\sigma_{0}=1.10$ (the dotted black curve) and $\sigma_{0}=0.29$ (the dashed blue curve) for $s=-1$. (b) $\sigma_{0}=0.40$ (the dotted black curve) and $\sigma_{0}=0.21$ (the dashed blue curve) for $s=1$. The solid red curves represent the corresponding intensity of the beam.}\label{fig1}
\end{figure}

However, the above two conditions are not sufficient physically for the existence of the HG solions.
The HG solitons exist only in the condition that the induced NRI is bell-shaped, or quasi-bell-shaped, because the solitary waves can be considered as the eigen-modes of the self-induced waveguide, i.e., the linear waveguide induced by the waves themselves~\cite{Snyder-ol-91}.
The bell-shaped NRI induced by the beam appears only within the rightmost range of the GDN obtained  above.
Taking the case of triple-solitons for example, the induced NRIs [given by Eq.~(\ref{index})] at different $\sigma_{0}$ are shown in Fig.~\ref{fig1}. For comparison, the corresponding intensity $I(x)[=|u_{2}(x)|^{2}$] of the HG wave [described by Eq.~(\ref{un})] are also shown.
We can see that for both $s=-1$ [Fig.~\ref{fig1} (a)] and $s=1$ [Fig.~\ref{fig1} (b)], the NRI with the $\sigma_{0}$ in the rightmost range, i.e., $\sigma_{0}\in(1.06, +\infty)$ and $\sigma_{0}\in (0.39, 0.79)$, respectively, are bell-shaped. Otherwise, the NRI with the $\sigma_{0}$ in other ranges, even close to the rightmost one, is not bell-shaped at all (the dashed blue curves in Fig.~\ref{fig1}), and can not construct a waveguide.
Likewise, in the cases of other $n$, the bell-shaped [or quasi-bell-shaped, as shown in Fig.~\ref{fig2} (a)] NRI appears only in the rightmost GDN ranges obtained from the minimal ``potential".
Following the same procedure, the ranges of $\sigma_{0}$ within which the HG solitons exist are given in Table \ref{tab-1} (the variational results).

 \begin{table}[htbp]
\centering
\caption{The ranges of $\sigma_{0}$ within which the HG-type solitons exist}\label{tab-1}
\begin{threeparttable}
\begin{tabular}{cc|c|p{2.0cm}<{\centering}|c|c}
  \hline \hline
  & &$n=0$\tnote{*}&$n=1$&$n=2$&$n\geq3$\\ \hline
  $s=-1$&variational&$(1.05, +\infty)$&$(1.06, +\infty)$ &$(1.06, +\infty)$&$(1.06, +\infty)$\\
  &numerical&$(1.05, +\infty)$&$(1.05, +\infty)$  &$(1.05, +\infty)$&$(1.05, +\infty)$\\ \hline
  $s=+1$&variational&$(0, 0.77)$&$(0.38, 0.79)$  &$(0.39, 0.79)$&$(0.39, 0.79)$\\
  &numerical&$(0.05, 0.78)$&$(0.38, 0.78)$\tnote{**}&$(0.39, 0.78)$&$(0.40, 0.78)$\\
  \hline \hline
\end{tabular}
\begin{tablenotes}
\centering
\footnotesize
\item[*] The cases that $n=0$ were discussed in the Ref.~\cite{Liang-njp-20}.
\item[**]It is the case referred to as the out-of-phase bound-state solitons in the Ref.~\cite{Liang-oe-16}, where the existence range of the soliton is $0.16\leq\sigma\leq0.78$, and the algorithm used is the imaginary-time method. However, the results on $\sigma<0.38$ in the Ref.~\cite{Liang-oe-16} are unreliable, because of the reason given in the Ref.~\cite{note}.
\end{tablenotes}
\end{threeparttable}
\end{table}

\section{Numerical solutions}\label{sec4}
For the nonlocal nonlinear system described by Eqs.~(\ref{se}) and (\ref{index}), we know from the variational analysis above that there exist the HG-type solitos on the conditions that $\sigma_{0}>1.06$ when $s=-1$, and $0.38<\sigma_{0}<0.79$ when $s=1$. Next we will search its multi-peak solitons by numerical methods according to the stationary equation
\begin{equation}\label{e-ss}
\frac{1}{2}\frac{\mathrm{d}^{2}u(x)}{\mathrm{d}x^{2}}+u(x)\int^{\infty}_{-\infty}R(x-\xi)|u(\xi)|^{2}\mathrm{d}\xi=bu(x),
\end{equation}
which is obtained by substituting Eq.~(\ref{ss}) into Eqs.~(\ref{se}) and (\ref{index}).
In the computation, the $w_{M}$ can be chosen arbitrarily, and a set of soliton solutions with any value of $w_{M}$ can be obtained from the solution of a given $w_{M}$, because the system has the transform invariance~\cite{Liang-oe-16}.

\begin{figure}[htbp]
  \centering
  \includegraphics[width=8.5cm]{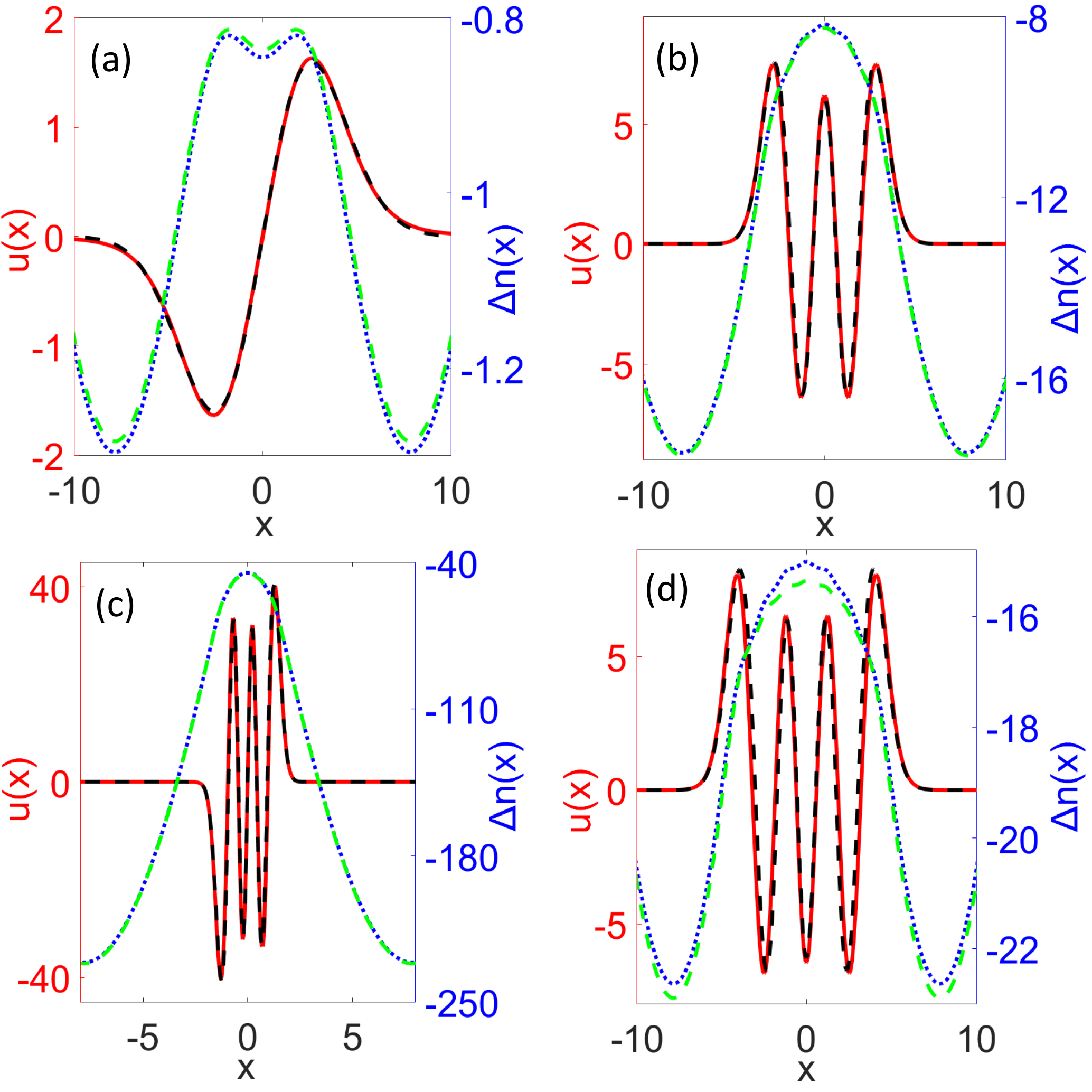}
   \caption{The profiles of the numerical multi-peak solitons $u(x)$ (the solid red curves) and the light induced NRI $\Delta n(x)$ (the dotted blue curves) for $s=-1$ and $w_{M}=5$. (a-d) for $n=1, 4, 5, 6$ and $\sigma_{0}=1.13, 1.43, 3.52, 1.08$, respectively. The variational results [the dashed black curves for $u(x)$, and the dashed green curves for $\Delta n(x)$] with the same parameters are also given for comparison.}\label{fig2}
\end{figure}

For the case that $s=-1$, the multi-peak solitons are obtained by the perturbation-iteration method~\cite{Hong-josab-18}, and the initial conditions are taken as the variational solutions. The numerical results show that the multi-peak solitons always exist when the initial GDN $\sigma_{0} >1.05$. The ranges of $\sigma_{0}$ obtained by the numerical method are also shown in Table \ref{tab-1}, from which we can observe that the numerical and variational values agree well with each other. The profiles $u(x)$ of several multi-peak solitons and the light induced NRIs $\Delta n(x)$ are shown in Fig.~\ref{fig2}, where $w_{M}=5$.
It can be found that the NRIs are of convex shapes at the center of the beams, although the values are negative.
In such cases, the beams can sample the self-focusing NRIs that construct the self-induced waveguides.
For comparison, the variational solutions with the same $\sigma_{0}$, $w_{M}$ and $n$ are also given. Clearly, the variational multi-peak solitons are in good agreement with the numerical ones.

\begin{figure}[htbp]
  \centering
  \includegraphics[width=8.5cm]{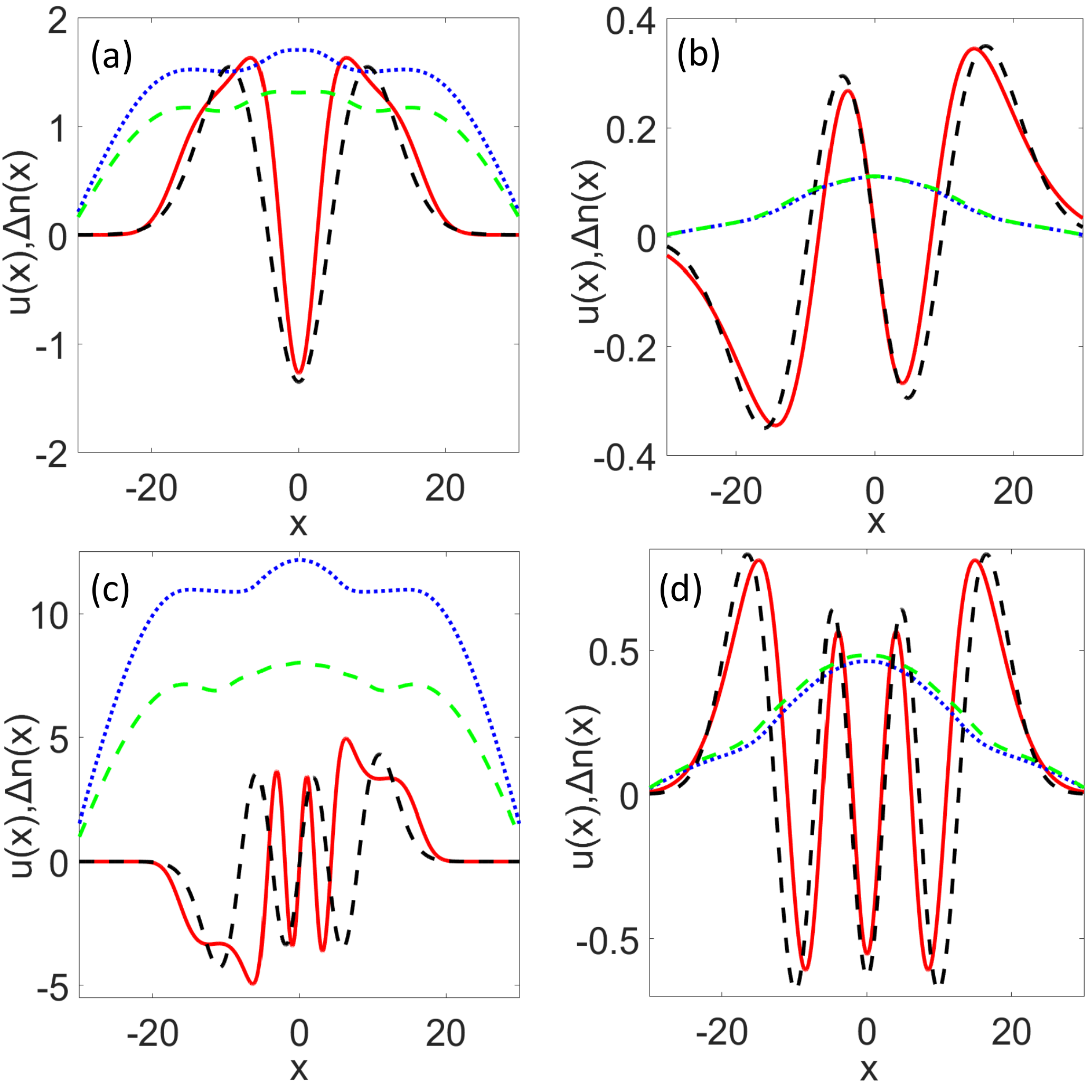}
  \caption{The profiles of the numerical multi-peak solitons $u(x)$ (the solid red curves) and the light induced NRI $\Delta n(x)$ (the dotted blue curves) for $s=1$ and $w_{M}=10$. (a-d) for $n=2, 3, 5, 6$ and $\sigma_{0}=0.74, 0.48, 0.75, 0.51$, respectively. The corresponding numerical results are denoted in the same way as in Fig.~\ref{fig2}. }\label{fig3}
\end{figure}

For the case that $s=1$, the numerical results are shown in Fig.~\ref{fig3}. We can find that the NRIs in the main area of the beams are almost positive, although they can be negative in the area with zero light intensity.
Like in the above case that $s=-1$, the corresponding variational HG solitons are shown for comparison. Obviously,
there appears more considerable difference between the numerical and variational ones compared with Fig.~\ref{fig2},
especially in Fig.~\ref{fig3}(a, c) with large $\sigma_{0}$.
The profiles $u(x)$ are HG-like when $\sigma_{0}$ is small [Fig.~\ref{fig3}(b, d)], while they deform to be with flat step-like wings and far from the HG-like shape when $\sigma_{0}$ becomes large [Fig.~\ref{fig3}(a, c)].
The GDN ranges within which the numerical multi-peak solitons exist are also given in Table \ref{tab-1}.
We can see that the numerical results agree well with the variational ones, too.
Here the multi-peak solitons with relatively small GDN (about $\sigma_{0}\leq0.5$) can be obtained by the perturbation-iteration method as in the case that $s=-1$. However, when $\sigma_{0}$ increases, the perturbation-iteration has a poor accuracy, even can not convergent (when $\sigma_{0}>0.65$). The reason may lies in that the perturbation-iteration method only applies to the solutions close to the HG function.
Thus, we should use another method, the Newton-conjugate-gradient method~\cite{Yang-jcp-09}, to obtain the soliton solutions on this condition.
Since the Newton-conjugate-gradient method is sensitive to the initial condition, that is, the initial condition should be reasonably close to the exact solution~\cite{Yang-jcp-09}, we at advance obtain the soliton with smaller GDN, by iterating with the solition acquired by the perturbation-iteration method as the initial condition. Then the solitons with larger GDN are iterated stepwise by taking the solitons with slightly smaller GDN as the initial conditions.

\begin{figure}[htbp]
  \centering
  \includegraphics[width=8.5cm]{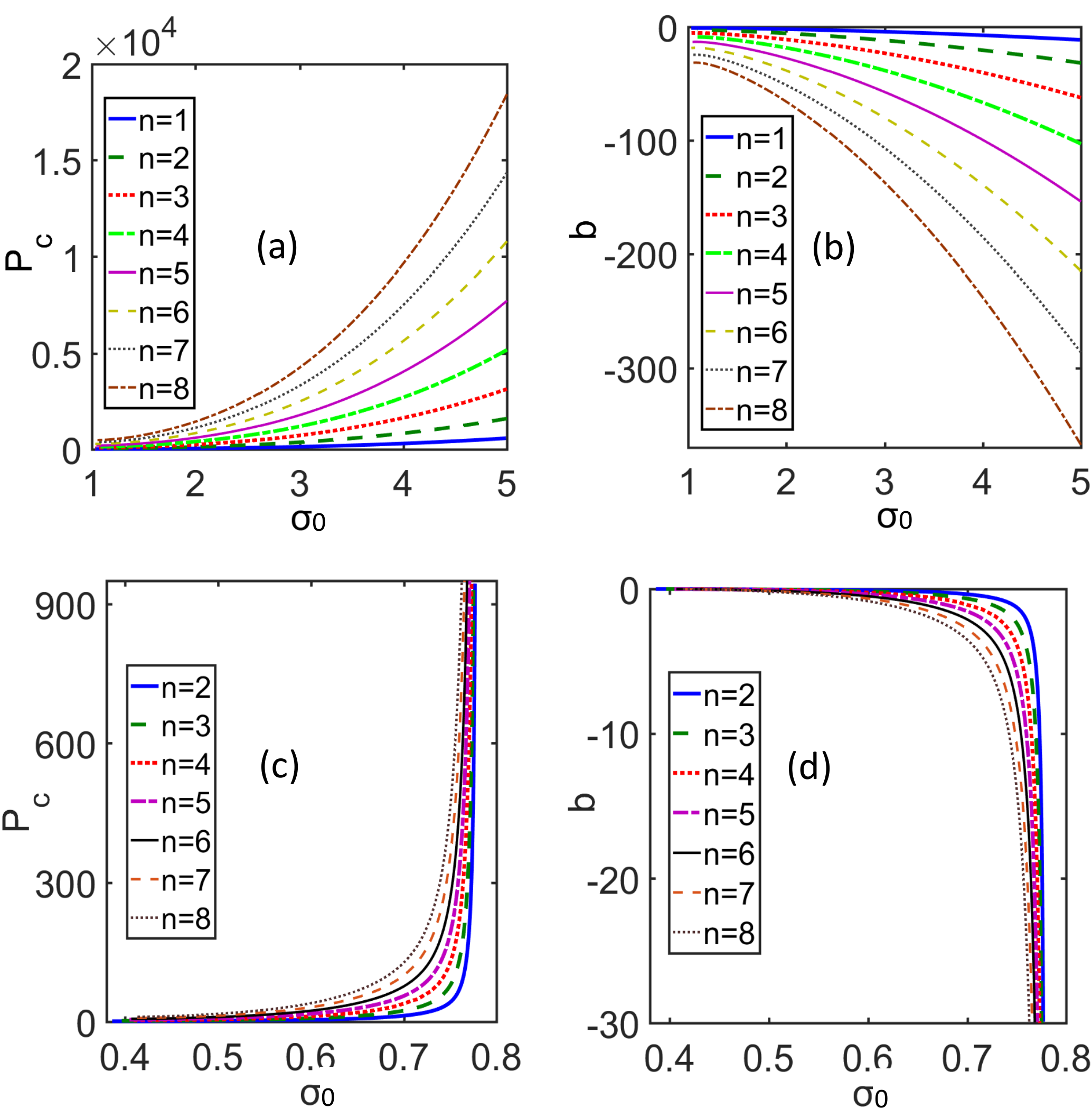}
  \caption{The dependencies of the power $P_{c}$ on the GDN $\sigma_{0}$ [(a, c)] and the propagation constant $b$ on the GDN $\sigma_{0}$ [ (b, d)] for the multi-peak solitons.  (a, b) and (c, d) for $s=-1$ and $s=1$, respectively.}\label{fig4}
\end{figure}

Then, let us investigate the dependencies of the power $P_{c}=\int_{-\infty}^{\infty}|u(x)|^{2}\mathrm{d}x$ and the propagation constant $b$ on the GDN $\sigma_{0}$ for the multi-peak solitons. We present the results with the peak numbers ranging from 2 ($n=1$) to 9 ($n=8$) in Fig.~\ref{fig4}, the upper row of which are for the case $s=-1$ with the GDN $\sigma_{0}\leq 5$, the bottom row of which are for the case $s=1$ with the GDN $\sigma_{0}$ in the whole existence range of the solitons.
Here the results about the dipole-solitons ($n=1$) for the case $s=1$ are omitted, since they, referred to as the out-of-phase solitons, had been discussed~\cite{Liang-oe-16}.
 It is clear from  Fig.~\ref{fig4} (a, c) that $P_{c}$ increase monotonically with $\sigma_{0}$ for all the multi-peak solitons. From Fig.~\ref{fig4} (b, d), we can see that all the propagation constants $b$ are negative, which decrease monotonically with $\sigma_{0}$. These results are similar to the low order solitons~\cite{Liang-pra-19, Liang-oe-16}, but the slope (the absolute value) of the curves increases with the peak number.

\section{Stability of the multi-peak solitons}\label{sec5}
Further more, we will study the stability of the HG-type solitons by the linear stability analysing~\cite{Xu-ol-05, Dong-pra-10, Liang-oe-16,Liang-pra-19}. The perturbed soliton solution is supposed to be of the form $q(x,z)=[u(x)+g(x,z)+ih(x,z)]\exp(ibz)$, where the real part $g(x,z)$ and the imaginary part $h(x,z)$ of the perturbation can grow with a complex rate $\delta$ on propagation. Linearization of Eq.~(\ref{se}) around the soliton solution $u(x)$ yields the eigenvalue equations~\cite{Liang-oe-16,Liang-pra-19}
\begin{equation}\label{eig1}
\delta g=-\frac{1}{2}\frac{d^{2}h}{dx^{2}}+bh-\Delta nh,
\end{equation}
\begin{equation}\label{eig2}
\delta h=-\frac{1}{2}\frac{d^{2}g}{dx^{2}}-bg+\Delta ng+u\Delta N,
\end{equation}
in which $\Delta N=2\int_{-\infty}^{\infty}R(x-x')u(x')g(x')dx'$ is the refractive index perturbation. By solving the system of Eqs.~(\ref{eig1}) and (\ref{eig2}) we obtain the eigenvalues $\delta$, a positive real part of which means the instability of the solution $u(x)$.

\begin{figure}[htbp]
  \centering
  \includegraphics[width=8.5cm]{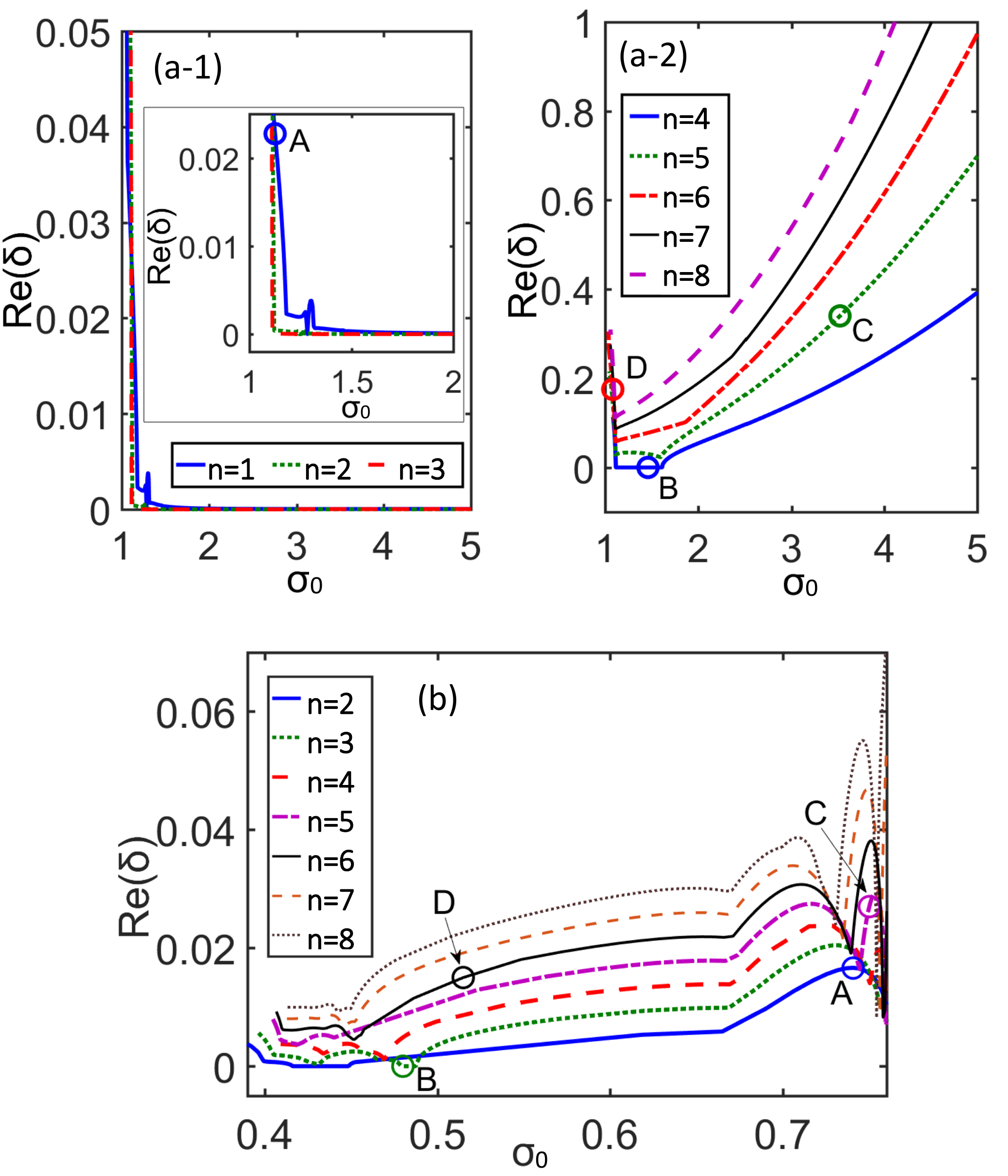}
   \caption{The maximal real parts of the perturbation growth rate $Re(\delta)$ versus the GDN $\sigma_{0}$ for the multi-peak solitons. (a) for  $s=-1$, the open circles denoted by  \textit{A} to \textit{D} corresponding to ($n=1, \sigma_{0}=1.13$), ($n=4, \sigma_{0}=1.43$), ($n=5, \sigma_{0}=3.52$) and ($n=6, \sigma_{0}=1.08$), respectively. (b) for  $s=1$, the open circles denoted by  \textit{A} to \textit{D} corresponding to ($n=2, \sigma_{0}=0.74$), ($n=3, \sigma_{0}=0.48$), ($n=5, \sigma_{0}=0.75$) and ($n=6, \sigma_{0}=0.51$), respectively.}\label{fig5}
\end{figure}

Fig.~\ref{fig5} shows the maximal real parts of the perturbation growth rate $Re(\delta)$ for the multi-peak solitons with different $n$, in which figures (a) and (b) are for the cases $s=-1$ and $s=1$, respectively. We can see from  Fig.~\ref{fig5} (a-1) that the dipole-, triple- and quadrupole-solitons ($n=1, 2, 3$, respectively) are stable
[$Re(\delta)=0$] when the GDN exceed certain values, while they are unstable [$Re(\delta)>0$] at small $\sigma_{0}$.
Fig.~\ref{fig5} (a-2) shows that the multi-peak solitons that $n>4$ are all unstable; however, when $n=4$ the (fifth-order) solitons can be stable
only within a small range of GDN. From figure (b), we can see that most of the $Re(\delta)$ are positive,
but $Re(\delta)=0$ for the triple- and quadrupole-solitons ($n=2, 3$, respectively) in very small ranges of GDN. That is, the multi-peak solitons are unstable on most conditions, except for the cases that $n=2,3$ within very small ranges of $\sigma_{0}$. The ranges of the GDN $\sigma_{0}$ within which the stable HG-type solitons exist are summarized in Table \ref{tab-2}.

\begin{table}[htbp]
\centering
\caption{The ranges of $\sigma_{0}$ within which the stable HG-type solitons exist}\label{tab-2}
\begin{threeparttable}
\begin{tabular}{p{1cm}|c|c}
  \hline \hline
   &$s=-1$&$s=+1$\\\hline
    $n=0$ \tnote{*}& $(1.05, +\infty)$&$(0.05, 0.78)$ \\
    $n=1$& $(1.41, +\infty)$&$(0.38, 0.78)$\tnote{**}\\
  $n=2$& $(1.10, +\infty)$&$(0.42, 0.45) $\\
  $n=3$& $(1.10, +\infty)$&$(0.48, 0.49)$ \\
  $n=4$& $(1.10, 1.61)$&no \\
  $n\geq5$&no&no\\
  \hline \hline
\end{tabular}
\begin{tablenotes}
\centering
\footnotesize
\item[*] The cases that $n=0$ were discussed in the Ref.~\cite{Liang-njp-20}.
\item[**] The value is different from the result in the Ref.~\cite{Liang-oe-16} because of the reason noted below Table \ref{tab-1} .
\end{tablenotes}
\end{threeparttable}
\end{table}

\begin{figure}[htbp]
  \centering
  \includegraphics[width=8.5cm]{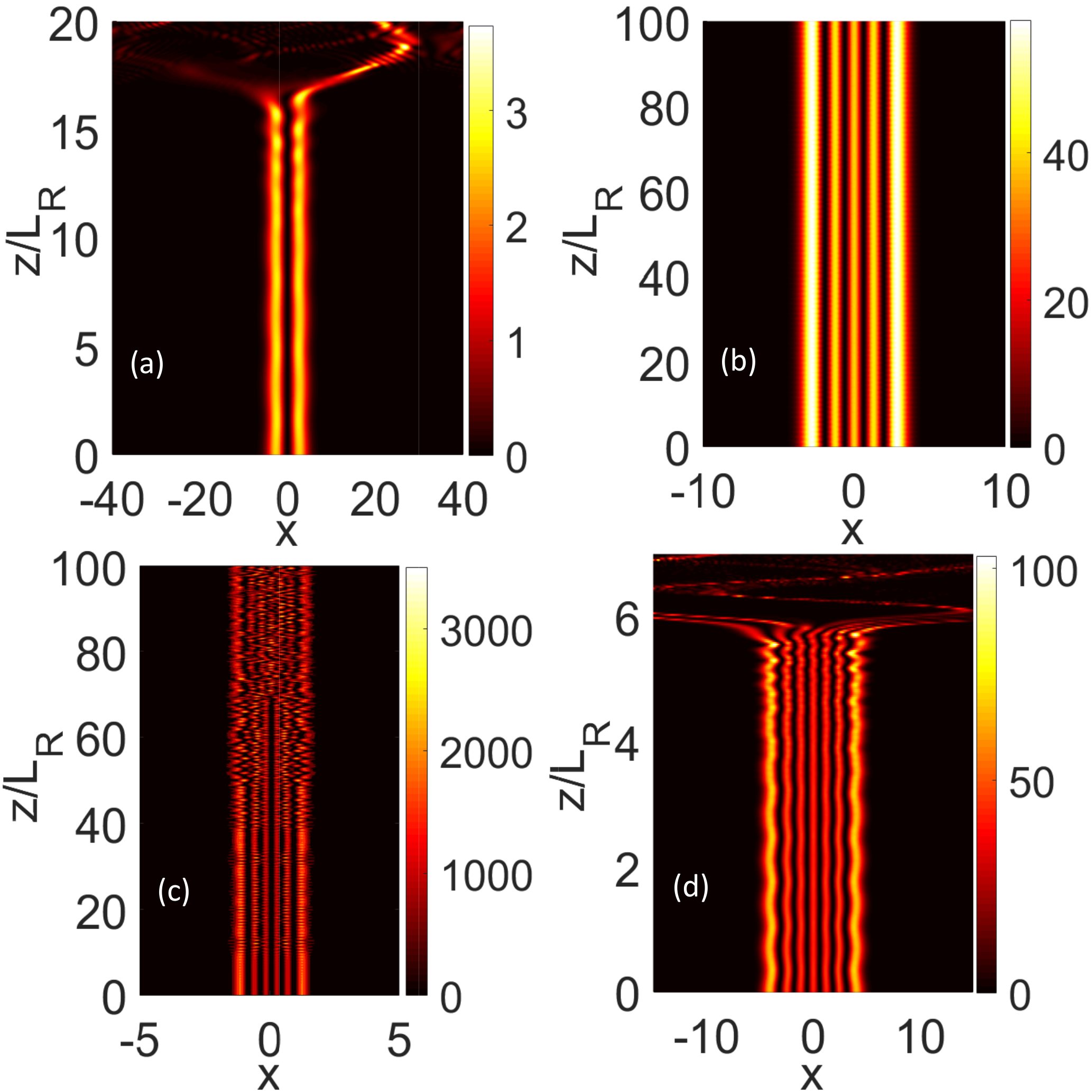}
   \caption{The contour plots of the intensity $I(x,z)$ for the propagation of several multi-peak solitons in the system that $s=-1$. (a-d) are corresponding to the cases denoted by \textit{A} to \textit{D} in Fig.\ref{fig5} (a), the initial inputs $u(x)$ of which are shown in Fig.~\ref{fig2} (a-d), respectively.}\label{fig6}
\end{figure}

\begin{figure}[htbp]
  \centering
 \includegraphics[width=8.5cm]{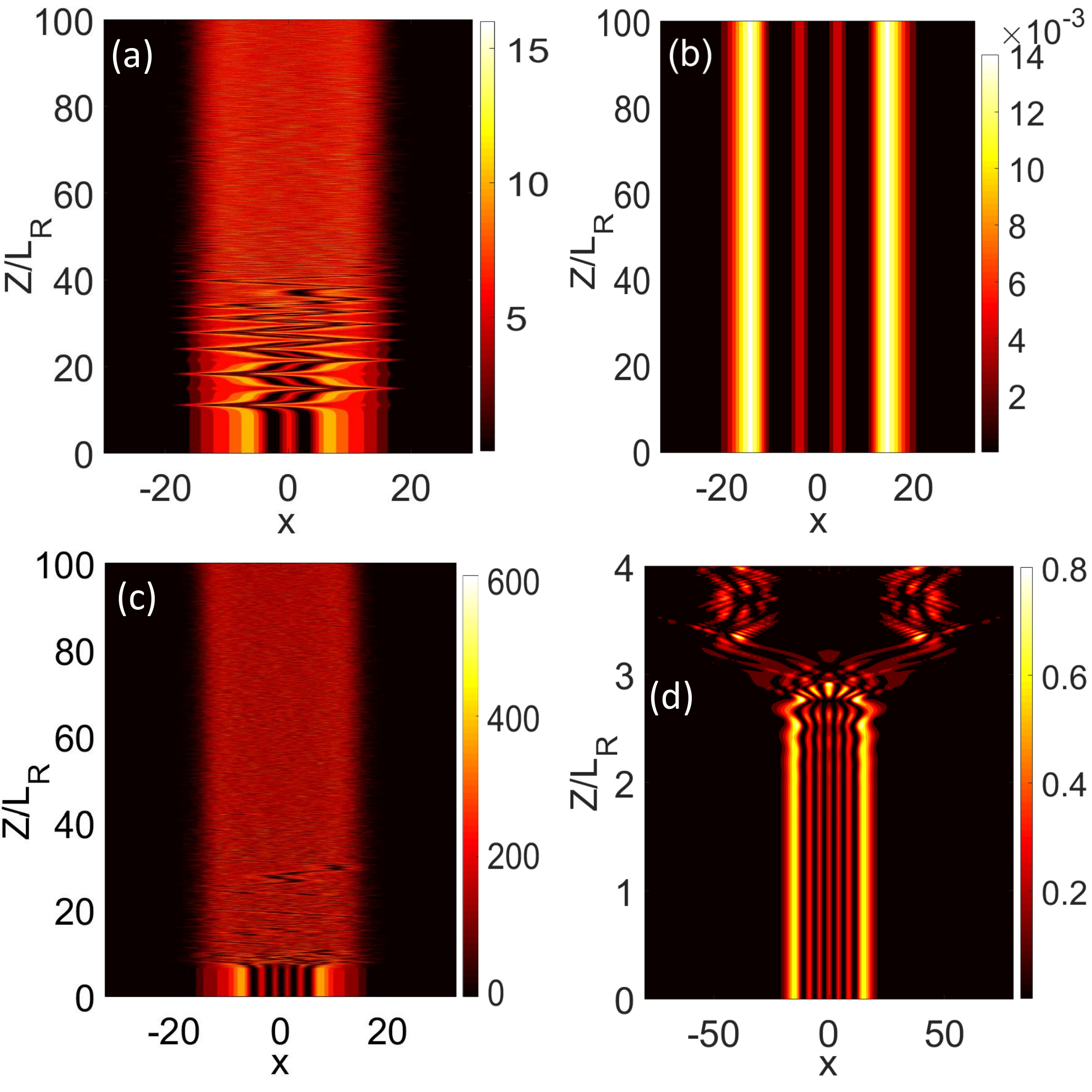}
   \caption{The contour plots of the intensity $I(x,z)$ for the propagation of several multi-peak solitons in the system that $s=1$.   (a-d) are corresponding to the cases denoted by \textit{A} to \textit{D} in Fig.\ref{fig5} (b), the initial inputs $u(x)$ of which are shown in Fig.~\ref{fig3} (a-d), respectively.}\label{fig7}
\end{figure}

 To confirm the above results of the stability analysis, we simulate the propagation of the multi-peak solitons with the input condition $q(x,0)=u(x)[1+\rho(x)]$, where $\rho(x)$ is a random function with a normal distribution and a standard deviation equal to 0.01. Fig.~\ref{fig6} depicts the contour plots of the intensity $I(x,z)$ [$=|q(x,z)|^2$ ] for the system that $s=-1$.  Fig.~\ref{fig6} (b) is for a stable fifth-order soliton, and the other three figures of  Fig.~\ref{fig6} are for unstable propagations of the multi-peak solitons.
 Similarly, the propagation of several multi-peak solitons for the case $s=1$ are shown in Fig.~\ref{fig7}.
Fig.~\ref{fig7} (b) is for the stable propagation of the quadrupole-soliton, the other three figures depict the unstable propagations of the solitons.
 It is clear that the stable propagations of the solitons agree well with the zero value of the $Re(\delta)$, while the unstable propagations of the solitons correspond to the positive $Re(\delta)$.  We can also find that, among the unstable cases, there are two quite different behaviours for the propagation of beams. In Fig.~\ref{fig6} (a, d) and  Fig.~\ref{fig7} (d), the beams break up after propagate several Rayleigh distances $L_{R}$ ($=w_{0}^{2}$); on the other hand, the beams keep self-trapped without spreading in Fig.~\ref{fig6} (c) and  Fig.~\ref{fig7} (a, c).  The former are for relatively small GDN, while the latter are for relatively large GDN, either $s=-1$ or $s=1$.

In the end of the paper, we would like to discuss the upper threshold for the peak number of the stable multi-peak solitons in different nonlocal nonlinear systems. The system  with the sine-oscillation response function can support the stable multi-peak solitons with at most five peaks.
On the other hand,  the systems with the exponential-decay response function~\cite{Xu-ol-05} and the logarithmic response function (in the lead glass)~\cite{Dong-pra-10} allow the stable multi-peak solitons with at most four peaks,
and the system with the Gaussian response function admits of no upper threshold for the number of peaks of stable solitons~\cite{Xu-ol-05, Deng-josab-07}.

\section{conclusions}\label{sec6}
In conclusion, the multi-peak solitons and their existence range for the nonlocal nonlinear system with the sine-oscillation response have been studied. By the variational approach, the HG solitons have been obtained, and their existence range have been acquired according to the minimal ``potential" and the bell-shaped NRI. To our knowledge, it is the first time to obtain analytically the existence range for the multi-peak solitons. The numerical multi-peak solitons have been obtained also, which agree with the variational ones, especially in the case of negative Kerr coefficient ($s=-1$). The linear stability analyses for the multi-peak solitons show that the system permits the stable solitons with at most five and four peaks for the cases $s=-1$ and $s=1$, respectively.

\section*{ACKNOWLEDGMENTS}
This research was supported by the National Natural Science Foundation of China (Grant No.~11704169) and the Science and Technology Program of Guangzhou (No. 2019050001).

\section*{Appendix A. Deviation of the evolution equation (\ref{width}) for the beam width}

The Euler-Lagrange equation corresponding to the variational problem
$\delta\int^{+\infty}_{0}L(A_{n},w_{n},c_{n},\alpha_{n},\dot{A}_{n},\dot{w}_{n},\dot{c}_{n},\dot{\alpha}_{n})\mathrm{d}z=0$ is
\begin{equation}\label{A-1}
\frac{\mathrm{d}}{\mathrm{d}z}(\frac{\partial L}{\partial\dot{q}_{i}})-\frac{\partial L}{\partial q_{i}}=0,  \tag{A.1}
\end{equation}
in which $L$ is the Lagrangian [Eq.~(\ref{lagran})],$q_{i}=\{A_{n},w_{n},c_{n},\alpha_{n}\}$, $\dot{q}_{i}=dq_{i}/dz$.
 Following the standard procedures of the variational approach, we get $\delta L/\delta A_{n}=0$, $\delta L/\delta w_{n}=0$, $\delta L/\delta c_{n}=0$, and $\delta L/\delta \alpha_{n}=0$. Then, we can obtain the four equations:
 \begin{equation}\label{A-a}
 -\frac{w^{3}_{n}A_{n}}{\sqrt{2n+1}}\left[\frac{\mathrm{d}c_{n}}{\mathrm{d}z}+\frac{(2n+1)^{2}}{2w^{4}_{n}}+2c_{n}^{2}\right]-\frac{2w_{n}A_{n}}{\sqrt{2n+1}}\frac{\mathrm{d}\alpha_{n}}{\mathrm{d}z}+\frac{2sA^{3}_{n}}{(\sqrt{\pi}2^{n}n!)^{2}}e_{n}=0, \tag{A.2}
 \end{equation}
 \begin{equation}\label{A-w}
 -\frac{3w^{2}_{n}A_{n}^{2}}{2\sqrt{2n+1}}\left[\frac{\mathrm{d}c_{n}}{\mathrm{d}z}+2c_{n}^{2}\right]+\frac{A^{2}_{n}(2n+1)^{3/2}}{4w^{2}_{n}}-\frac{A_{n}^{2}}{\sqrt{2n+1}}\frac{\mathrm{d}\alpha_{n}}{\mathrm{d}z}+\frac{sA^{4}_{n}}{2w_{n}(\sqrt{\pi}2^{n}n!)^{2}}f_{n}=0, \tag{A.3}
 \end{equation}
 \begin{equation}\label{A-c}
 -\frac{1}{2\sqrt{2n+1}}\frac{\mathrm{d}(w^{3}_{n}A_{n}^{2})}{\mathrm{d}z}+\frac{2w^{3}_{n}A_{n}^{2}}{\sqrt{2n+1}}c_{n}=0, \tag{A.4}
 \end{equation}
 \begin{equation}\label{A-alph}
 -\frac{1}{\sqrt{2n+1}}\frac{\mathrm{d}}{\mathrm{d}z}(w_{n}A_{n}^{2})=0. \tag{A.5}
 \end{equation}
The last equation (\ref{A-alph}) indicates that the power $P_{n}=w_{n}A_{n}^{2}/\sqrt{2n+1}$ is a constant.
By combining Eqs.~(\ref{A-a}) and (\ref{A-w}), we obtain
\begin{equation}\label{equ-22}
\frac{\mathrm{d}\alpha_{n}}{\mathrm{d}z}=-\frac{(2n+1)^{2}}{2w^{2}_{n}}+\frac{3sP_{n}(2n+1)}{4w_{n}^{2}(\sqrt{\pi}2^{n}n!)^{2}}[2e_{n}-f_{n}/3], \tag{A.6}
\end{equation}
By substituting the constant power, the Eq.~(\ref{A-c}) can lead to
\begin{equation}\label{equ-23}
\frac{\mathrm{d}c_{n}}{\mathrm{d}z}=-\frac{1}{2w^{2}_{n}}(\frac{\mathrm{d}w_{n}}{\mathrm{d}z})^{2}+\frac{1}{2w_{n}}\frac{\mathrm{d}^{2}w_{n}}{\mathrm{d}z^{2}}. \tag{A.7}
\end{equation}
Then we can obtain the second order differential equation (\ref{width}) about the evolution of the beam width by eliminating other variables from Eqs.~(\ref{A-c}), (\ref{equ-22}) and (\ref{equ-23}).


\begin{thebibliography}{99}
\bibitem{Assanto-book-13}
G. Assanto, Nematicons: Spatial Optical Solitons in Nematic Liquid Crystals (New Jersey: John Wiley \& Sons, 2013).
\bibitem{Guo-book-15}
Q. Guo, D. Lu, and D. Deng, Nonlocal spatial optical solitons, in: X. Chen, Q. Guo, W. She, H. Zhang, and G. Zhang (Eds.), Advances in Nonlinear Optics (Berlin: De Gruyter, 2015) pp. 227-305.
\bibitem{Snyder-science-97}
A.W. Snyder, and D.J. Mitchell, Accessible solitons, Science \textbf{276}, 1538 (1997).

\bibitem{Rasmussen-pre-05}
P.D. Rasmussen, O. Bang, and W. Krolikowski, Theory of nonlocal soliton interaction in nematic liquid crystals, Phys. Rev. E \textbf{72}, 066611 (2005).
\bibitem{Rotschild-np-06}
 C. Rotschild, B. Alfassi, O. Cohen, and M. Segev, Long-range interactions between optical solitons, Nat. Phys. \textbf{2}, 769 (2006)
 \bibitem{Hu-apl-06}
 W. Hu, T. Zhang, Q. Guo, L. Xuan, and S. Lan, Nonlocality-controlled interaction of spatial solitons in nematic liquid crystals, Appl. Phys. Lett. \textbf{89}, 071111 (2006) .

 \bibitem{Guo-pre-04}
Q. Guo, B. Luo, F. Yi, S. Chi, and Y. Xie,  Large phase shift of nonlocal optical spatial solitons,  Phys. Rev. E \textbf{69}, 016602(2004).
\bibitem{Shou-ol-11}
Q. Shou, X. Zhang, W. Hu, and Q. Guo,  Large phase shift of spatial solitons in lead glass,  Opt. Lett. \textbf{36}, 4194 (2011).
\bibitem{Shou-oc-15}
Q. Shou, M. Wu, and Q. Guo,  Large phase shift of (1+1)-dimensional nonlocal spatial solitons in lead glass,  Opt. Commun. \textbf{338}, 133 (2015).
 \bibitem{Rotschild-prl-05}
C. Rotschild, O. Cohen, O. Manela, and M. Segev,  Solitons in nonlinear media with an infinite range of nonlocality: first observation of coherent elliptic solitons and of vortex-ring solitons,  Phys. Rev. Lett. \textbf{95}, 213904 (2005).
\bibitem{Xu-ol-05}
Z. Xu, Y.V. Kartashov, and L. Torner,  Upper threshold for stability of multipole-mode solitons in nonlocal nonlinear media,  Opt. Lett. \textbf{30}, 3171 (2005).
\bibitem{Rotschild-ol-06}
 C. Rotschild, M. Segev, Z. Xu, Y.V. Kartashov, and L. Torner,  Two-dimensional multipole solitons in nonlocal nonlinear media,  Opt. Lett. \textbf{31}, 3312 (2006).
 \bibitem{Skupin-pre-06}
S. Skupin, O. Bang, D. Edmundson, and W. Krolikowski,  Stability of two-dimensional spatial solitons in nonlocal nonlinear media,  Phys. Rev. E \textbf{73}, 066603 (2006).
\bibitem{Buccoliero-prl-07}
D. Buccoliero, A.S. Desyatnikov, W. Krolikowski, and Y.S. Kivshar,  Laguerre and Hermite soliton clusters in nonlocal nonlinear media,  Phys. Rev. Lett. \textbf{98}, 053901 (2007).
\bibitem{Skupin-prl-07}
S. Skupin, M. Saffman, and W. Krolikowski,  Nonlocal stabilization of nonlinear beams in a self-focusing atomic vapor, Phys. Rev. Lett. \textbf{98}, 263902 (2007).
\bibitem{Deng-josab-07}
D. Deng,  X. Zhao, Q. Guo, and S. Lan,  Hermite-Gaussian breathers and solitons in strongly nonlocal nonlinear media, J. Opt. Soc. Am. B \textbf{24}, 2537 (2007).
\bibitem{Buccoliero-ol-08}
D. Buccoliero, A.S. Desyatnikov, W. Krolikowski, and Y.S. Kivshar,  Spiraling multivortex solitons in nonlocal nonlinear media, Opt. Lett. \textbf{33}, 198 (2008).
\bibitem{Rotschild-np-08}
 C. Rotschild, T. Schwartz, O. Cohen, and M. Segev,  Incoherent spatial solitons in effectively instantaneous nonlinear media,  Nat. Photon. \textbf{2}, 371 (2008).
 \bibitem{Dong-pra-10}
L. Dong, and F. Ye,  Stability of multipole-mode solitons in thermal nonlinear media, Phys. Rev. A \textbf{81}, 013815 (2010).
\bibitem{Song-oe-18}
L. Song, Z. Yang, X. Li and X. Zhang,  Controllable Gaussian-shaped soliton clusters in strongly nonlocal media,  Opt. Express \textbf{26}, 19182 (2018).
\bibitem{Zhong-sr-17}
L. Zhong, Y. Li,  Y. Chen, W. Hong, W. Hu, and Q. Guo,  Chaoticons described by nonlocal nonlinear Schr\"{o}dinger equation,  Sci. Rep. \textbf{7}, 41438 (2017).
\bibitem{Zhong-pra-19}
L. Zhong, Q. Guo, W. Hu, W. Hong, and W. Xie,  Chaotic self-trapped optical beams in strongly nonlocal nonlinear media, Phys. Rev. A \textbf{99}, 043816 (2019).
\bibitem{Shou-ol-09}
Q. Shou, Y. Liang, Q. Jiang, Y. Zheng, S. Lan, W. Hu, and Q.
Guo, Boundary force exerted on spatial solitons in cylindrical strongly nonlocal media, Opt. Lett. \textbf{34}, 3523 (2009).
\bibitem{Zeng-pra-18}
S. Zeng, M. Chen, T. Zhang, W. Hu, Q. Guo, and D. Lu, Analytical modeling of soliton interactions in a nonlocal nonlinear medium analogous to gravitational force, Phys. Rev. A \textbf{97}, 013817 (2018).
\bibitem{Ghofraniha-PRL-99}
N. Ghofraniha, C. Conti, G. Ruocco, and S. Trillo, Shocks in nonlocal media, Phys. Rev. Lett. \textbf{99}, 043903 (2007).
\bibitem{Litvak-jl-1966}
A.G. Litvak, Self-focusing of powerful light beams by thermal effects, JETP Lett. \textbf{4}, 230-232 (1966).
\bibitem{Conti-prl-04}
C. Conti, M. Peccianti, and G. Assanto, Observation of optical spatial solitons in a highly nonlocal medium, Phys. Rev. Lett. \textbf{92}, 113902 (2004).
\bibitem{Esbensen-pra-12}
B. K. Esbensen, M. Bache, W. Krolikowski, and O. Bang,  Quadratic solitons for negative effective secondharmonic diffraction as nonlocal solitons with periodic nonlocal response function,  Phys. Rev. A \textbf{86}, 023849 (2012).
\bibitem{Wang-ol-14}
J. Wang, Y. Li, Q. Guo, and W. Hu,  Stabilization of nonlocal solitons by boundary conditions,  Opt. Lett. \textbf{39}, 405 (2014).
\bibitem{Wang-oc-17}
Z. Wang, Q. Guo, W. Hong, and W. Hu,  Modulational instability in nonlocal Kerr media with sine-oscillatory response,  Opt. Commun. \textbf{394}, 31 (2017).
\bibitem{Chen-pra-18}
M. Chen, X. Ping, G. Liang, Q. Guo, D. Lu, and W. Hu,  Dark-bright quadratic solitons with a focusing effective Kerr nonlinearity, Phys. Rev. A \textbf{97}, 013829 (2018).
\bibitem{Liang-oe-16}
G. Liang, W. Hong, and Q. Guo,  Spatial solitons with complicated structure in nonlocal nonlinear media, Opt. Express \textbf{24}, 28790 (2016).
\bibitem{Liang-pra-19}
G. Liang, W. Hong, T. Luo, J. Wang, Y. Li, Q. Guo, W. Hu, and D.N. Christodoulides,  Transition between self-focusing and self-defocusing in a nonlocally nonlinear system,  Phys. Rev. A \textbf{99}, 063808 (2019).
\bibitem{Liang-njp-20}
G. Liang, D. Dang, W. Li, H. Li and Q. Guo, Nonlocality-controllable Kerr-nonlinearity in nonlocally nonlinear system
with oscillatory responses, New J. Phys. \textbf{22},073024 (2020).
\bibitem{Guan-sr-20}
J. Guan, Z. Ren, and Q. Guo, Stable solution of induced modulation instability, Sci. Rep. \textbf{10},10081 (2020).
\bibitem{Nikolov-pre-03}
N.I. Nikolov, D. Neshev, O. Bang, and W.Z. Krolikowski,  Quadratic solitons as nonlocal solitons,  Phys. Rev. E \textbf{68}, 036614 (2003).
\bibitem{Buryak-pla-95}
A.V. Buryak, Y.S. Kivshar, Solitons due to second harmonic generation, Phys. Lett. A \textbf{197}, 407-412 (1995).
\bibitem{Anderson-pra-83}
D. Anderson, Variational approach to nonlinear pulse propagation in optical fibers, Phys. Rev. A \textbf{27}, 3135-3145 (1983).
\bibitem{Zhong-oc-17}
L. Zhong, J. Yang, Z. Ren, and Q. Guo, Hermite Gaussian stationary solutions in strongly nonlocal nonlinear optical media, Opt. Commun. \textbf{383} 274-280 (2017).
\bibitem{Abramowitz-book-72}
M. Abramowitz and I.A. Stegun, Handbook of Mathematical Functions (New York: Dover, 1972 )  Chapter 22.
\bibitem{Snyder-ol-91}
A.W. Snyder, D.J. Mitchell, L. Poladian, and F. Ladouceur, Self-induced optical fibers: spatial solitary waves, Opt. Lett. \textbf{16} 21-23 (1991).
\bibitem{Hong-josab-18}
W. Hong, B. Tian, R. Li, Q. Guo and W. Hu, Perturbation-iteration method for multi-peak solitons in nonlocal nonlinear media,  J. Opt. Soc. Am. B \textbf{35}, 317 (2018).
\bibitem{Yang-jcp-09}
J. Yang, Newton-conjugate-gradient methods for solitary wave computation,  J. Comput. Phys. \textbf{228}, 7007 (2009).

\bibitem{note}
The intensities of the soliton solutions on $\sigma<0.38$ in the Ref.~\cite{Liang-oe-16} are found, later after the Ref.~\cite{Liang-oe-16} has been published, not to have enough zero zone near the two ends of the sample window, which is the same case for different windows. And the
soliton solutions on $\sigma>0.38$ in the Ref.~\cite{Liang-oe-16} coincide with the solutions obtained in this paper. There may be some problems to find the soliton solutins by the algorithm of the imaginary-time method when $\sigma<0.38$, as did in the Ref.~\cite{Liang-oe-16}.

\end{thebibliography}
\end{document}